\documentclass[prb,aps,draft,twocolumn,tightenlines,%
showpacs,floatfix]{revtex4}    
\usepackage{psfig}
\usepackage{amssymb}
\begin{document}

\title{Multi-subband  effect in spin dephasing in semiconductor 
quantum wells}
\author{M. Q. Weng}
\affiliation{Hefei National Laboratory for Physical Sciences at
  Microscale, University of Science and Technology of China,
Hefei, Anhui, 230026, China and\\
Department of Physics, University of Science and
Technology of China, Hefei, Anhui, 230026, China}%
\altaffiliation{Mailing Address}
\author{M. W. Wu}
\thanks{Author to whom correspondence should be addressed}%
\email{mwwu@ustc.edu.cn}%
\affiliation{Hefei National Laboratory for Physical Sciences at
  Microscale, University of Science and Technology of China,
Hefei, Anhui, 230026, China and\\
Department of Physics, University of Science and
Technology of China, Hefei, Anhui, 230026, China}%
\altaffiliation{Mailing Address}
\date{\today}

\begin{abstract}
Multi-subband effect on  spin precession and spin dephasing
in $n$-type GaAs quantum wells is studied with electron-electron and
electron-phonon scattering explicitly included. 
The effects of temperature, well width and applied electric field 
(in hot-electron regime)
on the spin kinetics are thoroughly investigated.  
It is shown that due to the strong inter-subband scattering, the spin
procession and the spin dephasing rate of electrons in different
subbands are almost identical despite the large difference in the
D'yakonov-Perel' (DP) terms of different subbands. It is also
shown that for quantum wells with small well width at temperatures
where only the lowest subband is occupied, the 
spin dephasing time increases with
the temperature as well as the applied in-plane electric field
 until the contribution from the second subband is
no longer negligible.  For wide quantum wells the spin dephasing time
tends to decrease with the temperature and the electric field.
\end{abstract}

\maketitle

Manipulating and transporting electron spins belong to the 
subjects of intense research in  semiconductor
spintronics,\cite{wolf,spintronics} which aims to incorporate the spin
degree of freedom into the traditional electronic devices. To date 
most of the technical proposals to manipulate the electron spin are
either through an applied magnetic field or via a gate voltage that
changes the spin-orbital coupling in confined semiconductor
structures through the Rashba effect.\cite{rashba,ras} 
Recently it has been realized that a strong in-plane electric field
also provides another path to
manipulate the spin dephasing in the presence of the spin-orbital coupling.
Rashba and Efros showed that an in-plane ac electric
field can manipulate electron spins efficiently.\cite{rashba_2003}
Pramanik {\em et al.} investigated the effect of an in-plane
electric field on the spin dephasing in GaAs quantum wire 
through Monte-Carlo simulation and revealed that the presence of the
strong electric field enhances the spin
 dephasing.\cite{pramanik_2003,pramanik_2003a}  A comprehensive 
many-body investigation
of the hot-electron effect on spin precession and spin
dephasing due to the strong in-plane electric field in $n$-type
GaAs quantum wells (QW's) has been reported recently\cite{weng_prb_2004b} 
in the electric-field and temperature regime where 
only the lowest subband is occupied.  
In that investigation all scattering, {\em ie.}, the
electron-electron,  electron-phonon and electron-impurity scattering
is explicitly included and is calculated self-consistently. It is
discovered however that the spin dephasing rate 
decreases with the applied electric field. 
 In the narrow QW, it is also discovered that the spin dephasing rate
 decreases with temperature.\cite{mali,weng_prb_2003}

The spin precession and spin 
 dephasing in wide QW's should show quite different 
temperature and electric-field dependence as
electrons populate more than one subband 
and therefore experience quite  different spin-orbital
 coupling strength.  
In the present paper, we study the multi-subband effect
on the spin precession/dephasing in $n$-type GaAs QW's,
where the spin dephasing mainly comes from the D'yakonov-Perel' (DP)
mechanism,\cite{dp}
with and without the high in-plane electric field.  
The inclusion of multi subbands allows us to 
investigate QW's with wide well width and also
allows us to study the regime of higher electric field where
the single subband model cannot deal with due to the 
``runaway effect''.\cite{dmitriev_2000,weng_prb_2004b}
We reveal the intra- and inter-subband scattering to the
spin dephasing  times of each subband. The model of our
investigation is composed of (100) GaAs QW's of width $a$
with its growth direction along the $z$-axis. An uniform electric
field $\mathbf{E}$ and a moderate magnetic field $\mathbf{B}$ are
applied along the $x$-axis (Vogit configuration).  Due to the
confinement of the QW, the momentum states along the $z$-axis are
quantized.  Therefore the electron states are characterized by a
subband index $n$ and a two-dimensional wavevector ${\bf k}=(k_x,
k_y)$, together with a spin index $\sigma$. 
With the DP term (DPT) included, the Hamiltonian of the electron
in the QW reads:
\begin{eqnarray}
  H&=&\sum_{n\sigma n^{\prime}\sigma^{\prime}\mathbf{k}}\bigl\{
  (\varepsilon_{n\mathbf{k}}-e\mathbf{E}\cdot\mathbf{R})
  \delta_{nn^{\prime}}
  \delta_{\sigma\sigma^{\prime}}
  +
  \bigl[g\mu_B\mathbf{B}+\mathbf{h}_{n n^{\prime}}
  (\mathbf{k})\bigr]\nonumber\\
 && \cdot
  {\mbox{\boldmath$\sigma$\unboldmath}_
  {\sigma\sigma^{\prime}}\over 2}\biggr\}
  c^{\dagger}_{n\mathbf{k}\sigma}
  c_{n^{\prime}\mathbf{k}\sigma^{\prime}}+H_I.
\label{eq:hamiltonian}
\end{eqnarray}
Here $\varepsilon_{n\mathbf{k}}=\mathbf{k}^2/2m^{\ast}
+\langle k_z^2\rangle_n/m^{\ast}$ is the energy
spectrum of the electron with momentum $\mathbf{k}$ and effective mass
$m^{\ast}$ in the $n$-th subband.   
\mbox{\boldmath$\sigma$\unboldmath} are the Pauli
matrices. $\mathbf{R}=(x,y)$ represents  the position.
$\mathbf{h}_n(\mathbf{k})$  is the DPT
which serves as an effective magnetic field
with its magnitude and direction depending
on ${\bf k}$. It is composed of
the Dresselhaus term\cite{dress} and the Rashba term.\cite{ras,rashba}
For GaAs QW, the leading term is
the Dresselhaus one which can be written as:
\begin{eqnarray}
  \label{eq:DP}
  h_{nn^{\prime},x}(\mathbf{k}) &=& \gamma 
  k_x (k_y^2-\langle n| k_z^2|n\rangle)
  \delta_{nn^{\prime}}
  \ ;\nonumber\\
  h_{nn^{\prime},y}(\mathbf{k}) &=& \gamma k_y 
  (\langle n|k_z^2|n\rangle-k_x^2)
  \delta_{nn^{\prime}}
  \ ;\nonumber\\
  h_{nn^{\prime},z}(\mathbf{k}) &=&  
  \gamma 
  \langle n |k_z|n^{\prime}\rangle
  (k_x^2-k_y^2)
  \ .
  \label{eq:dp}
\end{eqnarray}
In the above, $\langle n|k^2_z|n\rangle$ represents the average of the
operator $-({\partial\over\partial z})^2$ over the electronic state of
the $n$-th subband and is therefore $n^2(\pi/a)^2$ under the 
infinite-well-depth assumption and $\langle n|k_z|n^{\prime}\rangle$ 
is the matrix element of $-i\partial /\partial_z$ 
between $n$- and $n^{\prime}$-th
subbands. $\gamma$ here is the spin-orbital coupling constant.\cite{aronov} 
The interaction Hamiltonian $H_I$ is composed of the Coulomb
interaction $H_{ee}$, the electron-phonon scattering $H_{ph}$, as well
as the electron-impurity scattering $H_i$. Their expressions can be
found in textbooks.\cite{mahan,haug}

In order to study the hot-electron effect on spin dephasing, we limit
our system to a spacial homogeneous one in order to avoid the
additional complicity such as charge/spin diffusion. The kinetic Bloch
equations in such a system are constructed using the nonequilibrium
Green function method with the gradient expansion\cite{haug} and can
be written as: 
\begin{equation}
  \label{eq:Bloch}
  \dot{\rho}_{n\mathbf{k},\sigma\sigma^{\prime}} -
  e\mathbf{E}\cdot\nabla_{\mathbf{k}}
  \rho_{n\mathbf{k},\sigma\sigma^{\prime}}
  = \dot{\rho}_{n\mathbf{k},\sigma\sigma^{\prime}}|_{\mathtt{coh}}
  +\dot{\rho}_{n\mathbf{k},\sigma\sigma^{\prime}}|_{\mathtt{scatt}}\ ,
\end{equation}
where $\rho_{n\mathbf{k}\sigma\sigma^{\prime}}$ represent the single
particle density matrix elements. The diagonal terms describe the electron
distribution functions $\rho_{n\mathbf{k},\sigma\sigma}\equiv
f_{n\mathbf{k}\sigma}$. The off-diagonal elements
$\rho_{n\mathbf{k},{1\over 2}-{1\over 2}}=\rho_{n\mathbf{k},-{1\over
    2}{1\over 2}}^\ast
\equiv \rho_{n\mathbf{k}}$
stand for the inter-spin-band polarizations (spin  
coherence).\cite{wu_prb_2000} 
The second terms in the kinetic equations describe the
momentum and  energy input from the  electric field $\mathbf{E}$.
$\dot{\rho}_{n\mathbf{k}\sigma\sigma^{\prime}}|_{\mathtt{coh}}$
on the right hand side of the equations
describe the coherent spin precession around
the applied magnetic field ${\bf B}$,
the effective magnetic field ${\bf h}({\bf k})$
from the DPT as well as the effective magnetic field from the
electron-electron Coulomb interaction in the Hartree-Fock  order.
$\dot{\rho}_{n\mathbf{k}\sigma\sigma^{\prime}}|_{\mathtt{scatt}}$
denote the electron-impurity, the electron-phonons, as well
as the electron-electron Coulomb scattering. The electron-electron
Coulomb scattering comes from the Coulomb interaction beyond the
lowest order, {\em i.e.}, the Hartree-Fock order. 
It is noted that in writing the kinetic equations, we have neglected 
the inter-subband coherence
$\rho_{nn^{\prime}\mathbf{k}\sigma\sigma^{\prime}}
(n\not=n^{\prime})$ as these terms are much smaller than the intra-subband 
spin coherence due to the small ``pumping'' terms 
$h_{nn^{\prime},z}(\mathbf{k})$.

\begin{figure}[htbp]
  \vskip 1pc
  \centering
  \psfig{file=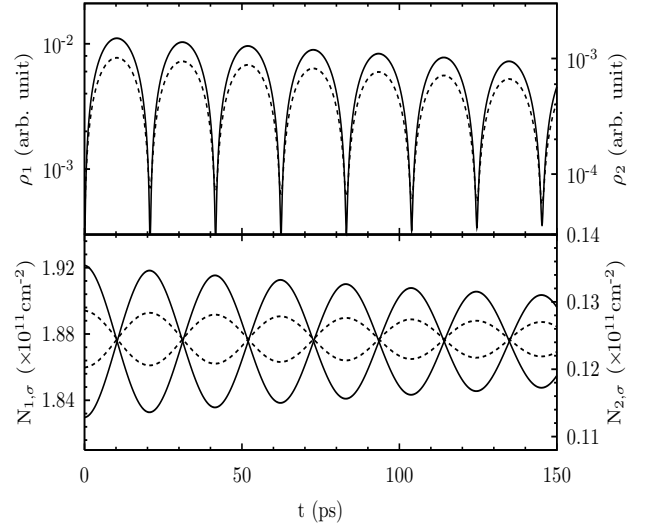,width=0.95\columnwidth,height=7cm}
  \caption{Temporal evolution of the spin signal
    with width $a=17.8$\ nm and electron density $N_e=4\times
    10^{11}$\ cm$^{-2}$ for $T=200$\ K and $B=4$\ T under zero applied
    electric field. The initial spin
    polarization $P=2.5$\ \%.
    Upper panel: The incoherently-summed  spin coherence
    $\rho_n$ of $n$-th subband;
    Lower panel: The electron densities of $n$-th subband with
    spin $\sigma$. The solid curves are for the first subband and the
    dashed ones are for the second subband. Note that the scales of the
    second subband are on the right-hand side of the figure.} 
  \label{fig:time}
\end{figure}

It is seen that all the unknowns to be solved
appear in the coherent and the scattering terms
nonlinearly. Therefore the kinetic Bloch equations have to be
solved self-consistently to obtain the electron distribution and the
spin coherence.
By numerically solving the kinetic Bloch equations in the 
self-consistent fashion, one is able to
obtain the temporal evolutions of the electron distribution functions
$f_{n\mathbf{k}\sigma}(t)$ and the spin coherence
$\rho_{n\mathbf{k}}(t)$.
Once these quantities are obtained,
all the quantities such as
electron mobility  $\mu_n$, hot-electron temperature $T_{e,n}$ 
 as well as the spin dephasing rate
 for electrons in each subband can be deduced. 
For subband $n$, the mobility is given by
$\mu_n=\sum_{\mathbf{k}\sigma}f_{n\mathbf{k}\sigma}\;k_x/[m^\ast
E\sum_{\mathbf{k}\sigma}f_{n\mathbf{k}\sigma}]$;
the electron temperature is obtained by fitting the Boltzmann
tail of the electron distribution function; whereas
the spin dephasing rate is determined by the slope of the envelope of the
incoherently summed spin coherence $\rho_n(t)=\sum_{\mathbf{k}}
|\rho_{n\mathbf{k}}(t)|$.\cite{kuhn_1992,wu_prb_2000,wu_js_2001}

The initial conditions at $t=0$ are taken to be
$\rho_{n\mathbf{k}}(0)=0$ and electron distribution functions 
are chosen to be those in the steady
state under the electric field but
without the magnetic field and the DPT.\cite{weng_prb_2004b}
Specifically $f_{n\mathbf{k},\sigma}(0)$ is the solution of the
kinetic equations (\ref{eq:Bloch}) in the steady state, with
the spin coherence $\rho_{n\mathbf{k}}$, the magnetic field  and the
DPT set to zero.
The implementation schemes of the numerical solution of the
kinetic equations and the initial
condition can be found in Ref.~\onlinecite{weng_prb_2004b}.
The total electron density $N_e$,
the applied magnetic field $B$ and the initial spin 
polarization $P=\sum_{n{\bf k}}(f_{n{\bf k}\frac{1}{2}}-
f_{n{\bf k}-\frac{1}{2}})/N_e$
are taken to be $4\times
10^{11}$\ cm$^{-2}$,  $4$\ T and 2.5\ \% respectively throughout the paper.

We first study the spin dephasing of electrons in a GaAs QW
with $a=17.8$\ nm. 
In this QW, the electrons
mainly distribute in the first two subbands under the applied electric
field up to 3\ kV/cm. Therefore, we only consider electrons 
in the first two subbands with $n$ being 1 and 2.

We first focus on the temporal evolution of the spin signals. In
Fig.~\ref{fig:time} we plot the densities of electrons with spin-up and
-down  as well as the corresponding incoherently-summed
spin coherence of each subband
versus time $t$ without applied electric field.
The background temperature $T=200$\ K.
 It is seen
from the figure that the electrons in both subbands undergo
damped oscillations with a same precession frequency and damping rate.
Nevertheless, 
as $\langle 2| k_z^2|2\rangle= 4\langle 1|k_z^2|1\rangle$, 
from Eq.\ (\ref{eq:DP}) electrons with a given wavevector in the second
subband experience an effective magnetic field from the DPT
with its magnitude {\em three times larger} than that of the first
subband.  At first glance one may expect that the
spin dephasing rate of electrons in the second
subband should be about 9 times faster than those
of electrons in the first subband.
It is  of particular interest to see that 
the decay rates of the spin signals of the 
two subbands are almost identical.

\begin{figure}[htbp]
  \centering
  \psfig{file=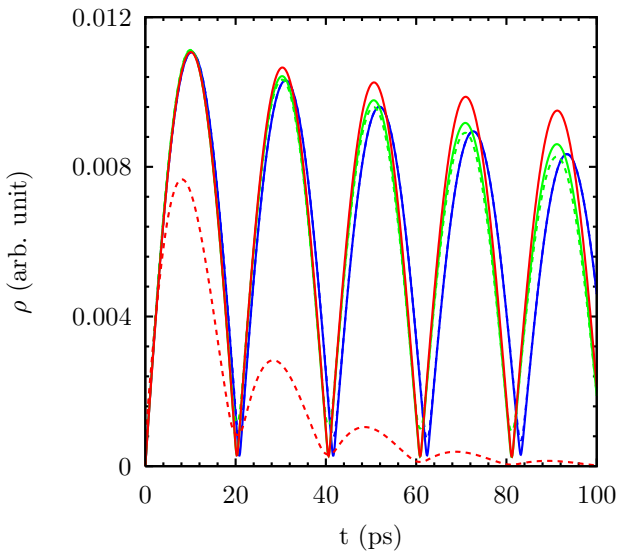,width=0.95\columnwidth}
  \caption{Temporal evolution of the incoherently-summed spin coherence of each
    subband $\rho_n$ for the same situation in Fig.\ 1. 
    The blue curves are $\rho_n(t)$ ($n=1$ and 2) from the full kinetic
    equations (Noted that $\rho_1(t)$ and $\rho_2(t)$
    match each other); The green curves are $\rho_n(t)$ without the
    inter-subband  Coulomb scattering; The red curves are $\rho_n(t)$ without
    the inter-subband  scattering. 
    Solid curves: $\rho_1(t)$; Dashed curves: $\rho_2(t)$. 
  }
  \label{fig:inters}
\end{figure}

In order to reveal the physics of the identicalness of the spin precession
and the dephasing of electrons in different subbands, we further
plot the temporal evolution of the spin coherence of the above case 
in Fig.~\ref{fig:inters}. For comparison we also
plot  $\rho_n(t)$ without the inter-subband
electron-electron Coulomb scattering (green curves) and those  
without the inter-subband electron-electron and electron-phonon
scattering, {\em i.e.}, without any inter-subband scattering
(red curves). It is seen from the figure that if we count for all
the scattering, the spin coherence of the different subbands
decays {\em identically}. However,
once the inter-subband of the Coulomb scattering is removed, the
  spin coherence of the lowest subband decays a little slower 
and that of the second subband decays a little faster than
that with all the scattering included.
If one further removes the inter-subband 
electron-phonon scattering, the spin coherence of the second subband 
decays very fast and disappears 
in the first 20\ ps while $\rho_1$ lasts much longer
than the case with all the inter-subband scattering included. It is
clear to see 
that as the inter-subband scattering is reduced, the identicalness 
of the spin  dephasing of electrons in different
subband is removed.  Once the inter-subband scattering is totally
removed, the spin coherence differs greatly between the different
subbands. Therefore, the identicalness is due to the strong
inter-subband scattering.

To further elucidate the effect of the inter-subband scattering 
on the spin dephasing, we adopt a much simplified model:
The system is simply described by the electron number $N_1$, $N_2$ and
the magnetic momentum $M_1$, $M_2$ of the two subbands. In the
non-degenerate case, the kinetics of the system are: 
$d N_i/dt = \sum_{j=1,2} \alpha_{ij} N_j$ and 
$d (N_iM_i)/dt =\sum_{j=1,2} \alpha_{ij}(N_jM_j) -
N_iM_i/\tau^s_{i}$  ($i=1$, 2),  
with  $\alpha_{ij}$ standing for the inter-subband scattering rate and
$1/\tau^s_i$ 
for the spin dephasing rate of the $i$-th subband. It is
easily seen that in the absence of the inter-subband scattering, the
spin signals in the different subbands decay with their own dephasing
rate. However, once a strong inter-subband scattering is presented
and the electron populations approach to the equilibrium, 
one is able to obtain that $\alpha_{ii} = -x_j/\tau$ and
$\alpha_{ij}=x_i/\tau$  ($j\not=i$) from the 
equilibrium condition $d N_i/dt=0$ and the 
detailed balance condition $\alpha_{ij}N_j=\alpha_{ji}N_i$ 
with $x_i=N_i/(N_1+N_2)$.  
$1/\tau$
is the inter-subband relaxation rate. 
Substituting these relations into the kinetic equations for the
magnetic moments $M_i$, one gets:
\begin{eqnarray}
  d M_1/d t &=& - x_2
  (M_1 - M_2)/\tau - M_1/\tau_1^s, \nonumber \\
  d M_2/d t &=&  x_1
  (M_1 - M_2)/\tau - M_2/\tau^s_2.
\end{eqnarray}
In the strong
inter-subband scattering limit, {\em i.e.}, $1/\tau\gg 1/\tau^s_i$, 
 the difference of magnetic momentum $M_1-M_2$ decays with the 
rate $1/\tau$.  That is the magnetic
momentums $M_i$ of different subbands  become almost identical to
each other in the time scale of $\tau$. Taking this fact into
account, one easily  gets the equation that controls the magnetic
momentum $M=M_1=M_2$: 
$d M/dt = -\sum_i (x_i/\tau_i^s) M$. Thus in the strong inter-subband
scattering limit, the magnetic momentums of different subbands
decay with the same rate 
$1/\tau_s=\sum_i x_i/\tau_i^s$. Therefore, it is understood that, 
in the presence of the strong inter-subband scattering, electrons
hop frequently among the subbands and fast exchange the spin signal. 
Consequently  electrons experience an average DP effective
magnetic field of the different subbands during the spin precession
and acquire the same  spin dephasing rate. Moreover, we also
discovered that in the case of strong electric field where the 
DPT can also change the spin precession
rates,\cite{weng_prb_2004b}  a strong inter-subband scattering also 
causes the identicalness of the spin precession rates of 
different subband, regardless from what expected from the DPT
analysis that the change of the spin precession rate of the 
second subband should be 3 times larger than that of the first one.

\begin{figure}[htbp]
  \vskip 1pc
  \centering
  \psfig{file=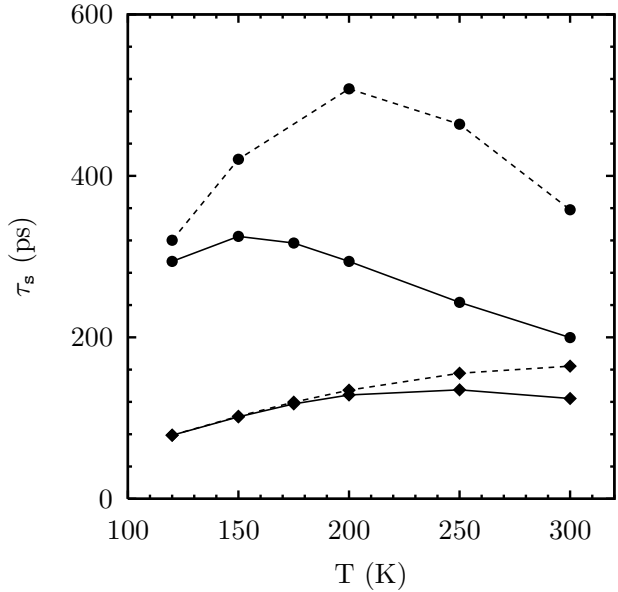,width=0.95\columnwidth}
  \caption{SDT $\tau_{s}$ {\em vs}. the background temperature
  $T$ without the applied electric field for two QW's with width
  $a=17.8$~nm ($\bullet$) and $12.7$~nm ($\blacklozenge$). The solid
  curves are the SDT calculated with the lowest two subbands included and
  the dashed curves are those calculated with only the lowest subband.  
  }
  \label{fig:tauT}
\end{figure}

We then turn to the temperature dependence of the spin dephasing of
electrons in QW's with different well width. 
The spin dephasing time (SDT)
versus the background temperature without applied electric field is
presented in Fig.~\ref{fig:tauT} for QW's with  $a=17.8$\ nm and 
$a=12.6$\ nm respectively. In the figure we also plot the SDT in the
single subband approximation as dashed curves in order to reveal
the contribution from the multi-subband effect. The figure shows that for
wide QW, the SDT first increases with the background temperature and then
decreases with it when the temperature rises to 150~K. However, for
narrow QW the SDT keeps increase until the temperature rises to
about $250$~K and then decreases slightly. 
The figure also shows the contribution of the higher subbands:
For wide QW, the population of electrons in the higher subbands is
larger even in the low temperature regime. As the temperature rises,
the difference of the SDT of multi-subband model and single-subband
one becomes larger and larger as the contribution of the higher
subbands becomes more important. While for the narrow QW, the contribution
of higher subbands is marginal when the temperature is lower than
250~K, thus the SDT's from the multi- and single-subband models are almost
the same in the regime $T<200$~K. After that, the SDT of
multi-subband model becomes smaller than that of single-subband one.
When the temperature rises higher than 250~K, the SDT of multi-subband
model decreases with $T$ while that of single-subband model keeps
growing upto the room temperature.

The different behavior of the temperature dependence of the SDT
 at different temperature regime and for different well width
originates from the DPT.
In $n$-type GaAs QW's the dominant spin dephasing 
mechanism is the spin dephasing caused by the inhomogeneous broadening 
induced by the DPT,
together with the spin conserving 
scattering.\cite{wu_epjb_2000,wu_js_2001} The
increase of the temperature brings many effects on the spin
dephasing through the DPT and the scattering rate: 
Firstly, the increase of temperature enhances the electron-phonon and the
electron-electron scattering, which 
tends  to drive electrons to a
more homogeneous state, and thus reduces the spin
dephasing.\cite{weng_prb_2003} 
Nevertheless, the increase of the temperature also enhances
the effect from the DPT. 
One can see from Eq.~(\ref{eq:dp}) that the DPT 
in the QW are composed of the terms linear 
and cubic in wave vector
$\mathbf{k}=(k_x,k_y,0)$.  The effects from the both terms increase
with  temperature as electrons are driven to larger
wave-vector states. This tends to bring a faster SDT.
 
For narrow QW's, the linear term is the dominate one even
for the lowest subband as $\langle k^2_z\rangle$ is much larger
than $k^2$ in the electron density and temperature regime of our
investigation. 
Although the effect of the linear term increases with 
temperature, the increase rate is slower than that of   
of the scattering. Therefore in the single-subband model 
the SDT of the narrow QW keeps increasing with temperature. 
However, for wide QW's, the linear term
of the lowest subband and the cubic term of the DPT
are comparable. When the temperature rises to 200\ K, the contribution of
cubic term becomes larger than that of the 
linear one of the lowest subband for the QW with $a=17.8$\ nm. As the 
effect of the cubic term
increases much faster with temperature than the scattering, 
the SDT decreases with temperature in the regime
where the effect of the cubic term
takes over that 
of the linear one. 

When the contribution
of the higher subbands are included, the increase of the temperature
further enhances the DPT through exciting  electrons to the
higher subbands which brings the much larger DPT. As a result, 
for the QW with $a=17.8$\ nm, the SDT
increases only in the regime $T<150$\ K and then decreases
when the effect of the multi-subband is
included. Similarly, the SDT of the 12.7\ nm QW  no longer
increases monotonically  with the temperature but first increases with
temperature and then decreases near room temperature. 

\begin{figure}[htbp]
  \centering
  \psfig{file=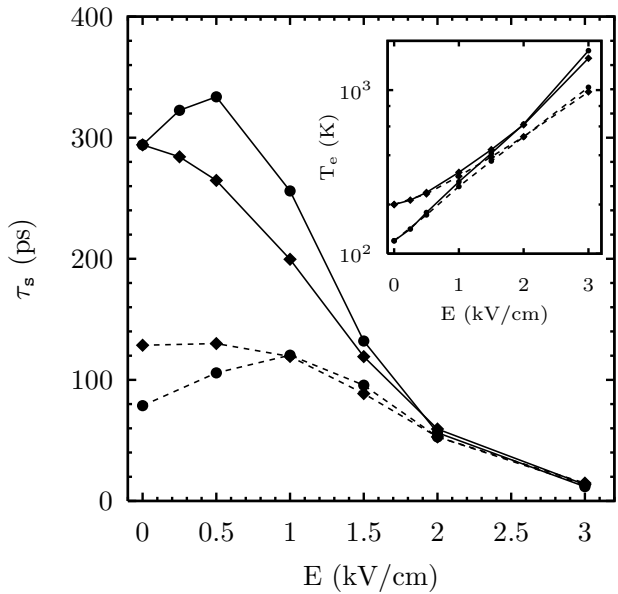,width=0.95\columnwidth}
  \caption{SDT {\em vs}. the applied electric
    field $E$ at different temperatures and  well widths:
    $\bullet$, $T=120$~K;  $\blacklozenge$, $T=200$~K; 
Solid curves, $a=17.8$~nm; Dashed
 curves, $a=12.7$~nm. Inset: The corresponding
    electron temperature $T_e$ as a function of the electric field.}
  \label{fig:tauE}
\end{figure}

We now study the effect of the applied electric field on the spin
precession and spin dephasing. In Fig.~\ref{fig:tauE}, we plot the SDT
as a function of the applied electric field for two temperatures and
two well widths. It is seen from the figure that for low temperature,
the SDT's first increase with $E$  for both wells
just as the case in the single-subband model.\cite{weng_prb_2004b} 
With further increase of the electric field, $\tau_s$ of 
the wide QW decreases rapidly
with $E$ once it is larger than 0.5\ kV/cm, while that
of the narrow QW remains growing until $E$ is higher than
1\ kV/cm. For high temperature, the SDT decreases with
$E$ even in small electric field regime for the wide QW. While 
it is insensitive to the applied electric field in the regime
$E<1$\ kV/cm for the narrow QW.

It is understood that the electric field can affect the spin dephasing
in two competing ways:\cite{weng_prb_2004b} On one hand, when the
electric field increases, the scattering rate increases due to the
hot-electron effect.  Consequently the electrons are driven to a more
homogeneous state in $k$-space and thus the spin dephasing is 
reduced---Effect I; 
On the other hand, the applied electric field also drives the electrons
to higher momentum states or higher subbands by rising the electron
temperature  and the drifting velocity and thus the spin
dephasing is enhanced through the strengthening of the 
DPT---Effect II.  
In the low temperature and low applied electric field regime, where
the linear DPT  is more important
and the electron 
drift velocity is small, the enhancement of the DP effect is
relatively smaller and the reduction of the inhomogeneous broadening
due to the hot-electron effect is more pronounced. Therefore the SDT
is raised in this regime. 
However, when the
electric field becomes larger, more electrons are driven to high
momentum states where the cubic DPT dominates,
moreover electrons are easier to be excited to
the higher subbands where the linear DPT
is much larger than that of the lowest subband, the second effect 
becomes more important and the SDT decreases with $E$ consequently. 

For wide QW's, electrons are much easier to be excited to higher
subbands and the domination of the 
cubic DPT is much  
easier to achieve than narrow QW's. Therefore at low
temperature the SDT of the wide
QW starts to decrease at relatively  smaller electric field (0.5\ kV/cm in
the figure). While the SDT of narrow QW
keeps increasing until the electric field is 1\ kV/cm, which is
about two times as large as that of the  wide well.
At high temperature, as the contribution from the 
cubic DPT is more important and also more electrons are excited to 
high subbands, Effect II is more
important even at low electric field for wide QW's. As 
a result the SDT of wide QW decreases monotonically  with the electric field. 
For the narrow QW in our calculation, both effects
of the applied electric field is comparable in the regime $E<1$\ kV/cm,
consequently the SDT is insensitive to the electric field in this
regime. 

It is also interesting to note that at very high electric 
field regime ($E\ge 2$~kV/cm), the SDT
is only determined by the electric field and is
insensitive to the well width and the background temperature.
 In this regime,
electrons are excited to high energy states by the electric field
and the electron temperature $T_e$ is much higher than the
background temperature $T$ as shown in the inset of
Fig.~\ref{fig:tauE}. In this situation, the major scattering mechanism
is the electron-electron Coulomb scattering and the dominant term of the DP
term is the cubic one. Both rely on the
electron temperature and are insensitive to the well width and the
background temperature in the high electric field regime. 
Nevertheless it is  noted that 
notwithstanding the fact that the electric field goes to very high in 
Fig.\ \ref{fig:tauE}, the inter-valley scattering between $\Gamma$-$L$ valleys, which
plays important role in the 
high-field transport, is not included. Once this additional scattering is
incorporated, the electron temperature will be  reduced and the spin
dephasing time may differ from what is predicted
in Fig.\ \ref{fig:tauE} when $E\ge 2$\ kV/cm. However a detailed study
of this inter-valley effect is beyond the
 scope of this paper and also 
the spin dephasing mechanism in the $L$ (and also $X$) valley 
is not full understood to date.

In conclusion, we have studied the multi-subband effect on spin
precession and spin dephasing in $n$-type GaAs QW's where 
electrons may occupy more than one subband. Our results show that due
to the strong inter-subband scattering,  electrons  
hop frequently among the subbands and thus experience an average DP effective
magnetic field regardless 
which subband the electrons are located. Therefore the spin
precession and the spin dephasing of electrons in each subband are almost
identical even though the DPT's in  different subbands vary
greatly. In the system we study, the spin dephasing is determined by
the joint effects of the inhomogeneous broadening induced by the DPT
and the spin-conserving scattering such as the electron-phonon and the
electron-electron Coulomb scattering. 
The scattering tends to randomize electrons in the momentum
space and reduces the inhomogeneous broadening and
consequently leads to the rise of the
SDT. Whereas the enhancement of the contribution of the DPT 
boosts the inhomogeneous broadening and therefore reduces the SDT.
When the distribution of electrons varies due to the
change of the well width, the background temperature and the
applied electric field, the contributions from the 
linear and the cubic DPT's and the contribution from the 
higher subbands as well as
the contribution from the spin conserving scattering change accordingly
and the SDT varies consequently. 

For narrow QW where electrons occupy only the lowest subband 
when the temperature is not too high,
 the dominant contribution from the DPT is the 
linear term. Therefore the SDT increases with  temperature
almost upto the room temperature as a result of the fast increase in
the scattering rate and the relatively 
slower increase of effect from the linear  DPT.
While for wide QW, the contributions from the 
cubic  DPT and
the higher subbands are more important, the SDT only slightly increases
with temperature in low temperature regime and then decreases with it. 
Similar effect also appears in the applied electric field dependence of
the SDT. In low temperature and low electric field regime, the SDT 
increases with
the electric field as the hot-electron effect effectively reduces the
inhomogeneous broadening of the DPT. When the background temperature
and/or the applied electric field increase, the SDT decreases as the
contributions from the higher momentum states and high subbands
become important. Whereas for wide QW's, as the contributions from the 
cubic DPT and the high subbands are more important, the SDT begins to
decrease at smaller electric field than that of narrow QW's. 

This work was supported by the Natural Science Foundation of China
under Grant No. 90303012.  MWW was also supported by  
the ``100 Person Project'' of Chinese Academy of
Sciences and the Natural Science Foundation of China under Grant No.
10247002. MQW was partially supported by China
Postdoctoral Science Foundation.


\begin{thebibliography}{20}
\expandafter\ifx\csname natexlab\endcsname\relax\def\natexlab#1{#1}\fi
\expandafter\ifx\csname bibnamefont\endcsname\relax
  \def\bibnamefont#1{#1}\fi
\expandafter\ifx\csname bibfnamefont\endcsname\relax
  \def\bibfnamefont#1{#1}\fi
\expandafter\ifx\csname citenamefont\endcsname\relax
  \def\citenamefont#1{#1}\fi
\expandafter\ifx\csname url\endcsname\relax
  \def\url#1{\texttt{#1}}\fi
\expandafter\ifx\csname urlprefix\endcsname\relax\def\urlprefix{URL }\fi
\providecommand{\bibinfo}[2]{#2}
\providecommand{\eprint}[2][]{\url{#2}}

\bibitem[{\citenamefont{Wolf}(2000)}]{wolf}
\bibinfo{author}{\bibfnamefont{S.~A.} \bibnamefont{Wolf}}, \bibinfo{journal}{J.
  Supercond.: Incorping Novel Magnetism} \textbf{\bibinfo{volume}{13}},
  \bibinfo{pages}{195} (\bibinfo{year}{2000}).

\bibitem[{\citenamefont{Ziese and Thornton}(2001)}]{spintronics}
\bibinfo{editor}{\bibfnamefont{M.}~\bibnamefont{Ziese}} \bibnamefont{and}
  \bibinfo{editor}{\bibfnamefont{M.~J.} \bibnamefont{Thornton}}, eds.,
  \emph{\bibinfo{title}{Spin Electronics}} (\bibinfo{publisher}{Springer},
  \bibinfo{address}{Berlin}, \bibinfo{year}{2001}).

\bibitem[{\citenamefont{Bychkov and Rashba}(1984{\natexlab{a}})}]{rashba}
\bibinfo{author}{\bibfnamefont{Y.~A.} \bibnamefont{Bychkov}} \bibnamefont{and}
  \bibinfo{author}{\bibfnamefont{E.~I.} \bibnamefont{Rashba}},
  \bibinfo{journal}{JETP Lett.} \textbf{\bibinfo{volume}{39}},
  \bibinfo{pages}{78} (\bibinfo{year}{1984}{\natexlab{a}}).

\bibitem[{\citenamefont{Bychkov and Rashba}(1984{\natexlab{b}})}]{ras}
\bibinfo{author}{\bibfnamefont{Y.~A.} \bibnamefont{Bychkov}} \bibnamefont{and}
  \bibinfo{author}{\bibfnamefont{E.~I.} \bibnamefont{Rashba}},
  \bibinfo{journal}{J. Phys. C} \textbf{\bibinfo{volume}{17}},
  \bibinfo{pages}{6039} (\bibinfo{year}{1984}{\natexlab{b}}).

\bibitem[{\citenamefont{Rashba and Efros}(2003)}]{rashba_2003}
\bibinfo{author}{\bibfnamefont{E.~I.} \bibnamefont{Rashba}} \bibnamefont{and}
  \bibinfo{author}{\bibfnamefont{A.~L.} \bibnamefont{Efros}},
  \bibinfo{journal}{Appl. Phys. Lett.} \textbf{\bibinfo{volume}{83}},
  \bibinfo{pages}{5295} (\bibinfo{year}{2003}).

\bibitem[{\citenamefont{Pramanik et~al.}(2003)\citenamefont{Pramanik,
  Bandyopadhyay, and Cahay}}]{pramanik_2003}
\bibinfo{author}{\bibfnamefont{S.}~\bibnamefont{Pramanik}},
  \bibinfo{author}{\bibfnamefont{S.}~\bibnamefont{Bandyopadhyay}},
  \bibnamefont{and} \bibinfo{author}{\bibfnamefont{M.}~\bibnamefont{Cahay}},
  \bibinfo{journal}{Phys. Rev. B} \textbf{\bibinfo{volume}{68}},
  \bibinfo{pages}{075313} (\bibinfo{year}{2003}).

\bibitem[{\citenamefont{Pramanik et~al.}(2004)\citenamefont{Pramanik,
  Bandyopadhyay, and Cahay}}]{pramanik_2003a}
\bibinfo{author}{\bibfnamefont{S.}~\bibnamefont{Pramanik}},
  \bibinfo{author}{\bibfnamefont{S.}~\bibnamefont{Bandyopadhyay}},
  \bibnamefont{and} \bibinfo{author}{\bibfnamefont{M.}~\bibnamefont{Cahay}},
  \bibinfo{journal}{Appl. Phys. Lett.} \textbf{\bibinfo{volume}{84}},
  \bibinfo{pages}{266} (\bibinfo{year}{2004}).

\bibitem[{\citenamefont{Weng et~al.}(2004)\citenamefont{Weng, Wu, and
  Jiang}}]{weng_prb_2004b}
\bibinfo{author}{\bibfnamefont{M.~Q.} \bibnamefont{Weng}},
  \bibinfo{author}{\bibfnamefont{M.~W.} \bibnamefont{Wu}}, \bibnamefont{and}
  \bibinfo{author}{\bibfnamefont{L.}~\bibnamefont{Jiang}},
  \bibinfo{journal}{Phys. Rev. B} \textbf{\bibinfo{volume}{69}}
  (\bibinfo{year}{2004}), \bibinfo{note}{in Press}.

\bibitem[{\citenamefont{Malinowski et~al.}(2000)\citenamefont{Malinowski,
  Britton, Grevatt, Harley, Ritchie, and Simmons}}]{mali}
\bibinfo{author}{\bibfnamefont{A.}~\bibnamefont{Malinowski}},
  \bibinfo{author}{\bibfnamefont{R.~S.} \bibnamefont{Britton}},
  \bibinfo{author}{\bibfnamefont{T.}~\bibnamefont{Grevatt}},
  \bibinfo{author}{\bibfnamefont{R.~T.} \bibnamefont{Harley}},
  \bibinfo{author}{\bibfnamefont{D.~A.} \bibnamefont{Ritchie}},
  \bibnamefont{and} \bibinfo{author}{\bibfnamefont{M.~Y.}
  \bibnamefont{Simmons}}, \bibinfo{journal}{Phys. Rev. B}
  \textbf{\bibinfo{volume}{62}}, \bibinfo{pages}{13034} (\bibinfo{year}{2000}).

\bibitem[{\citenamefont{Weng and Wu}(2003)}]{weng_prb_2003}
\bibinfo{author}{\bibfnamefont{M.~Q.} \bibnamefont{Weng}} \bibnamefont{and}
  \bibinfo{author}{\bibfnamefont{M.~W.} \bibnamefont{Wu}},
  \bibinfo{journal}{Phys. Rev. B} \textbf{\bibinfo{volume}{68}},
  \bibinfo{pages}{075312} (\bibinfo{year}{2003}).

\bibitem[{\citenamefont{D'yakonov and Perel'}(1971)}]{dp}
\bibinfo{author}{\bibfnamefont{M.~I.} \bibnamefont{D'yakonov}}
  \bibnamefont{and} \bibinfo{author}{\bibfnamefont{V.~I.}
  \bibnamefont{Perel'}}, \bibinfo{journal}{Zh. Eksp. Teor. Fiz.}
  \textbf{\bibinfo{volume}{60}}, \bibinfo{pages}{1954} (\bibinfo{year}{1971}),
  \bibinfo{note}{[Sov. Phys.-JETP {\bf 33}, 1053 (1971)]}.

\bibitem[{\citenamefont{Dmitriev et~al.}(2000)\citenamefont{Dmitriev,
  Kachorovskii, Shur, and Stroscio}}]{dmitriev_2000}
\bibinfo{author}{\bibfnamefont{A.~P.} \bibnamefont{Dmitriev}},
  \bibinfo{author}{\bibfnamefont{V.~Y.} \bibnamefont{Kachorovskii}},
  \bibinfo{author}{\bibfnamefont{M.~S.} \bibnamefont{Shur}}, \bibnamefont{and}
  \bibinfo{author}{\bibfnamefont{M.}~\bibnamefont{Stroscio}},
  \bibinfo{journal}{Solid State Commun.} \textbf{\bibinfo{volume}{113}},
  \bibinfo{pages}{565} (\bibinfo{year}{2000}).

\bibitem[{\citenamefont{Dresselhaus}(1955)}]{dress}
\bibinfo{author}{\bibfnamefont{G.}~\bibnamefont{Dresselhaus}},
  \bibinfo{journal}{Phys. Rev.} \textbf{\bibinfo{volume}{100}},
  \bibinfo{pages}{580} (\bibinfo{year}{1955}).

\bibitem[{\citenamefont{Aronov et~al.}(1983)\citenamefont{Aronov, Pikus, and
  Titkov}}]{aronov}
\bibinfo{author}{\bibfnamefont{A.~G.} \bibnamefont{Aronov}},
  \bibinfo{author}{\bibfnamefont{G.~E.} \bibnamefont{Pikus}}, \bibnamefont{and}
  \bibinfo{author}{\bibfnamefont{A.~N.} \bibnamefont{Titkov}},
  \bibinfo{journal}{Zh. Eksp. Teor. Fiz.} \textbf{\bibinfo{volume}{84}},
  \bibinfo{pages}{1170} (\bibinfo{year}{1983}), \bibinfo{note}{[Sov. Phys.-JETP
  {\bf 57}, 680 (1983)]}.

\bibitem[{\citenamefont{Mahan}(1981)}]{mahan}
\bibinfo{author}{\bibfnamefont{G.~D.} \bibnamefont{Mahan}},
  \emph{\bibinfo{title}{Many-particle Physics}} (\bibinfo{publisher}{Plenum},
  \bibinfo{address}{New York}, \bibinfo{year}{1981}).

\bibitem[{\citenamefont{Haug and Jauho}(1996)}]{haug}
\bibinfo{author}{\bibfnamefont{H.}~\bibnamefont{Haug}} \bibnamefont{and}
  \bibinfo{author}{\bibfnamefont{A.~P.} \bibnamefont{Jauho}},
  \emph{\bibinfo{title}{Quantum Kinetics in Transport and Optics of
  Semiconductors}} (\bibinfo{publisher}{Springer-Verlag},
  \bibinfo{address}{Berlin}, \bibinfo{year}{1996}).

\bibitem[{\citenamefont{Wu and Metiu}(2000)}]{wu_prb_2000}
\bibinfo{author}{\bibfnamefont{M.~W.} \bibnamefont{Wu}} \bibnamefont{and}
  \bibinfo{author}{\bibfnamefont{H.}~\bibnamefont{Metiu}},
  \bibinfo{journal}{Phys. Rev. B} \textbf{\bibinfo{volume}{61}},
  \bibinfo{pages}{2945} (\bibinfo{year}{2000}).

\bibitem[{\citenamefont{Kuhn and Rossi}(1992)}]{kuhn_1992}
\bibinfo{author}{\bibfnamefont{T.}~\bibnamefont{Kuhn}} \bibnamefont{and}
  \bibinfo{author}{\bibfnamefont{F.}~\bibnamefont{Rossi}},
  \bibinfo{journal}{Phys. Rev. Lett.} \textbf{\bibinfo{volume}{69}},
  \bibinfo{pages}{977} (\bibinfo{year}{1992}).

\bibitem[{\citenamefont{Wu}(2001)}]{wu_js_2001}
\bibinfo{author}{\bibfnamefont{M.~W.} \bibnamefont{Wu}}, \bibinfo{journal}{J.
  Supercond.: Incorping Novel Magnetism} \textbf{\bibinfo{volume}{14}},
  \bibinfo{pages}{245} (\bibinfo{year}{2001}),
  \bibinfo{note}{cond-mat/0109258}.

\bibitem[{\citenamefont{Wu and Ning}(2000)}]{wu_epjb_2000}
\bibinfo{author}{\bibfnamefont{M.~W.} \bibnamefont{Wu}} \bibnamefont{and}
  \bibinfo{author}{\bibfnamefont{C.~Z.} \bibnamefont{Ning}},
  \bibinfo{journal}{Eur. Phys. J. B.} \textbf{\bibinfo{volume}{18}},
  \bibinfo{pages}{373} (\bibinfo{year}{2000}).

\end{thebibliography}
\end{document}